\documentclass[aps, prd, floatfix, nofootinbib, superscriptaddress, twocolumn]{revtex4-2}

\usepackage{latexsym}
\usepackage{amsmath}
\usepackage{amssymb}
\usepackage{amsfonts}
\usepackage{textcomp}
\usepackage{color}
\usepackage{CJKutf8}

\usepackage[mathscr,scaled=1.15]{urwchancal}
\DeclareFontFamily{OT1}{pzc}{}
\DeclareFontShape{OT1}{pzc}{m}{it}%
{<-> s * [1.15] pzcmi7t}{}
\DeclareMathAlphabet{\mathpzc}{OT1}{pzc}{m}{it}

\usepackage{color}

\usepackage{supertabular}
\usepackage{placeins}
\usepackage{epsfig}
\usepackage{graphicx}

\definecolor{purple}{rgb}{0.5,0,0.5}
\definecolor{blue}{rgb}{0.0,0,0.9}
\definecolor{prdblue}{rgb}{0.133,0.118,0.498}
\usepackage[colorlinks=true, pdfstartview=FitV, linkcolor=prdblue, citecolor= prdblue, urlcolor=prdblue]{hyperref}

\makeatletter

\setbox0\hbox{$\xdef\scriptratio{\strip@pt\dimexpr
    \numexpr(\sf@size*65536)/\f@size sp}$}

\newcommand{\scriptveryshortarrow}[1][3pt]{{%
    \hbox{\rule[\scriptratio\dimexpr\fontdimen22\textfont2-.2pt\relax]
               {\scriptratio\dimexpr#1\relax}{\scriptratio\dimexpr.4pt\relax}}%
   \mkern-4mu\hbox{\let\f@size\sf@size\usefont{U}{lasy}{m}{n}\symbol{41}}}}

\makeatother

\begin{document}

\begin{CJK}{UTF8}{song}

\title{$\,$\\[-6ex]\hspace*{\fill}{\normalsize{\sf\emph{Preprint no}.\ NJU-INP 075/23}}\\[1ex]%
All-Orders Evolution of Parton Distributions: Principle, Practice, and Predictions}

\date{2023 June 05}

\author{Pei-Lin Yin
        $\,^{\href{https://orcid.org/0000-0001-7198-8157}{\textcolor[rgb]{0.00,1.00,0.00}{\sf ID}}}$}
\affiliation{College of Science, Nanjing University of Posts and Telecommunications, Nanjing 210023, China}

\author{Yin-Zhen.~Xu
       $^{\href{https://orcid.org/0000-0003-1623-3004}{\textcolor[rgb]{0.00,1.00,0.00}{\sf ID}},}$}
\affiliation{Dpto.~Ciencias Integradas, Centro de Estudios Avanzados en Fis., Mat. y Comp., Fac.~Ciencias Experimentales, Universidad de Huelva, Huelva 21071, Spain}
\affiliation{Dpto.~Sistemas F\'isicos, Qu\'imicos y Naturales, Univ.\ Pablo de Olavide, E-41013 Sevilla, Spain}

\author{\\Zhu-Fang~Cui
    $\,^ {\href{https://orcid.org/0000-0003-3890-0242}{\textcolor[rgb]{0.00,1.00,0.00}{\sf ID}}}$}
\affiliation{School of Physics, Nanjing University, Nanjing, Jiangsu 210093, China}
\affiliation{Institute for Nonperturbative Physics, Nanjing University, Nanjing, Jiangsu 210093, China}

\author{Craig D.~Roberts%
       $^{\href{https://orcid.org/0000-0002-2937-1361}{\textcolor[rgb]{0.00,1.00,0.00}{\sf ID}},}$}
\affiliation{School of Physics, Nanjing University, Nanjing, Jiangsu 210093, China}
\affiliation{Institute for Nonperturbative Physics, Nanjing University, Nanjing, Jiangsu 210093, China}

\author{Jos\'e Rodr\'iguez-Quintero%
       $^{\href{https://orcid.org/0000-0002-1651-5717}{\textcolor[rgb]{0.00,1.00,0.00}{\sf ID}},}$}
\affiliation{Dpto.~Ciencias Integradas, Centro de Estudios Avanzados en Fis., Mat. y Comp., Fac.~Ciencias Experimentales, Universidad de Huelva, Huelva 21071, Spain}

\begin{abstract}
\vspace*{-3ex}

\begin{small}
\centerline{
\href{mailto:yinpl@njupt.edu.cn}{yinpl@njupt.edu.cn} (PLY);
\href{mailto:yinzhen.xu@dci.uhu.es}{yinzhen.xu@dci.uhu.es} (YZX);}
\centerline{\href{mailto:phycui@nju.edu.cn}{phycui@nju.edu.cn} (ZFC);
\href{mailto:cdroberts@nju.edu.cn}{cdroberts@nju.edu.cn} (CDR);
\href{mailto:ose.rodriguez@dfaie.uhu.es}{jose.rodriguez@dfaie.uhu.es} (JRQ)
}
\end{small}
\vspace*{1ex}

Parton distribution functions (DFs) are defining expressions of hadron structure.
Exploiting the role of effective charges in quantum chromodynamics, an algebraic scheme is described which, given any hadron's valence parton DFs at the hadron scale, delivers predictions for all its DFs -- unpolarised and polarised -- at any higher scale.
The scheme delivers results that are largely independent of both the value of the hadron scale and the pointwise form of the charge; and,
%
%
\emph{inter alia}, enables derivation of a model-independent identity that relates the strength of the proton's gluon helicity DF, $\Delta G_p^\zeta$, to that of the analogous singlet polarised quark DF and valence quark momentum fraction.
Using available data fits and theory predictions, the identity yields
$\Delta G_p(\zeta_{\rm C}=\surd 3{\rm GeV})=1.48(10)$.
It furthermore entails that the measurable quark helicity contribution to the proton spin is $\tilde a_{0p}^{\zeta_{\rm C}}=0.32(3)$, thereby reconciling contemporary experiment and theory.
\end{abstract}

\maketitle

\end{CJK}

\begin{figure}[b]
\centerline{%
\includegraphics[clip, width=0.40\textwidth]{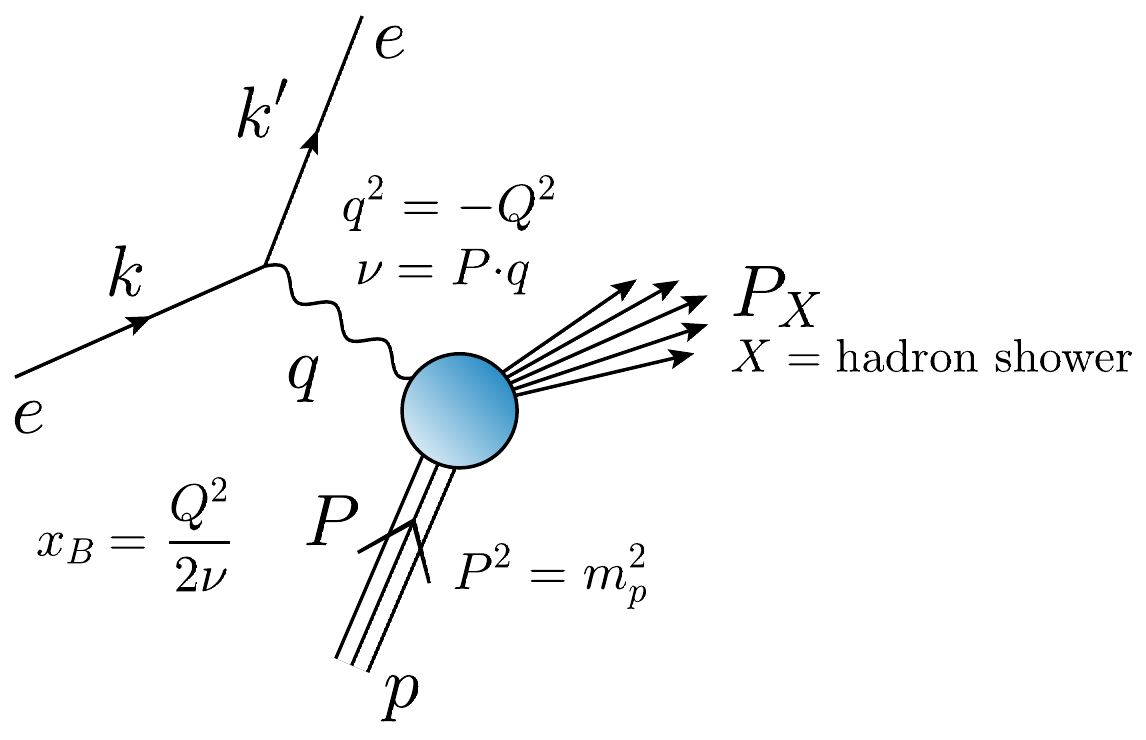}}
\caption{\label{Fdis}
electron + proton ($ep$) deep inelastic scattering: the initial-state proton disintegrates following collision with the virtual photon, producing a cascade, $X$, of final state hadrons.  With $m_p$ being the proton mass, the Bjorken limit is defined by the following kinematic conditions: $Q^2 \gg m_p^2$, $\nu\gg m_p^2$, $x_B\,=$\,fixed ($0<x_B<1$).
}
\end{figure}

\noindent\emph{1.$\;$Introduction}.
The Standard Model of particle physics has three components, one of which -- quantum chromodynamics (QCD) -- is supposed to describe strong interactions, \emph{i.e}., the interactions between gluons and quarks that lead to the formation of mesons, baryons and, ultimately, nuclei.  QCD originated with development of the quark model almost sixty years ago \cite{GellMann:1964nj, Zweig:1981pd} and crystalised following discovery of pointlike constituents of the proton (quarks) in electron + proton deep inelastic scattering (DIS) experiments \cite{Taylor:1991ew, Kendall:1991np, Friedman:1991nq, Friedman:1991ip} -- see Fig.\,\ref{Fdis}.

\begin{figure}[b]
\centerline{%
\includegraphics[clip, width=0.23\textwidth]{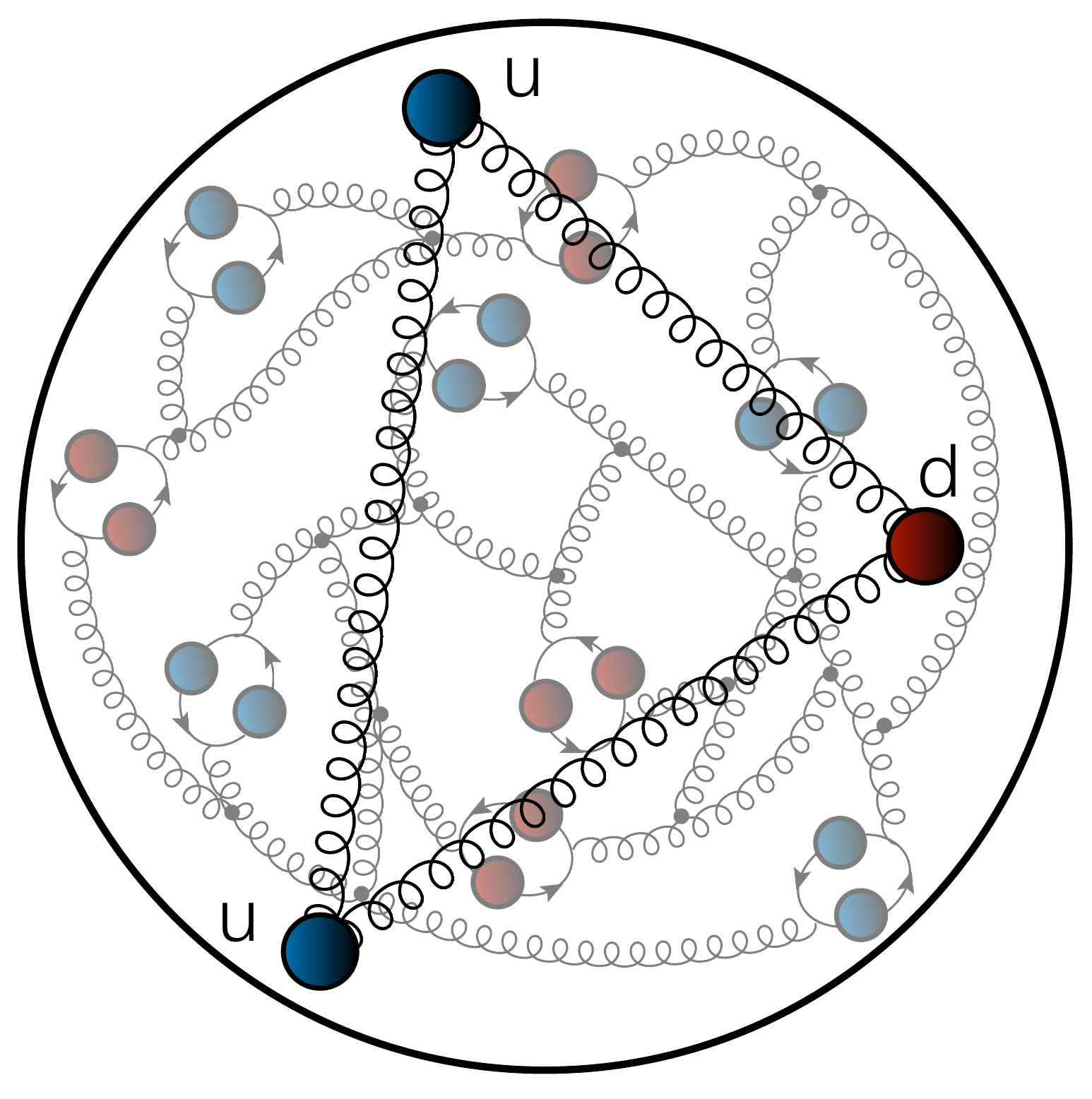}}
\caption{\label{ImageProton}
In terms of the fields used to express QCD's Lagrangian density, a proton contains two valence up ($u$) quarks and one valence down ($d$) quark; along with infinitely many gluons and sea quarks, drawn as ``springs'' and closed loops.
}
\end{figure}

Since the electron probe is well understood, such DIS experiments are seen as a measurement of the target proton's structure.  The formalism is now textbook material, \emph{e.g}., Ref.\,\cite{Ellis:1991qj}-[Ch.\,4].  In the Bjorken limit -- see Fig.\,\ref{Fdis}, the proton's two structure functions become equivalent and the measurement can be interpreted in terms of quark parton distribution functions (DFs) \cite{Feynman:1973xc}: writing $x\approx x_B$, then ${\mathpzc q}(x) dx$ is the probability that a quark within the proton carries a light-front fraction of the proton's momentum which lies in the range $[x,x+dx]$, $x\in[0,1]$.
Owing to the character of quantum field theory, as illustrated by Fig.\,\ref{ImageProton}, DFs associated with glue and sea-quark degrees of freedom (d$^{\rm s}$of) also play a role in the QCD description of scattering processes.
DFs are amongst the most fundamental properties of the target.  They have been a principal focus of experiment and theory for fifty years and the potential for new, deeper insights is growing with the operation of upgraded and construction of next-generation facilities \cite{Denisov:2018unj, Aguilar:2019teb, Brodsky:2020vco, Chen:2020ijn, Anderle:2021wcy, Arrington:2021biu, Aoki:2021cqa, Quintans:2022utc}.

Crucially, parton DFs are independent of the measurement process.  In fact, they are related to the modulus-squared of the target hadron light-front wave function \cite{Brodsky:1979gy}.  Notably, an analysis of the DIS process in the infinite momentum frame, where the hadron is considered to have momentum $P\approx (P,0,0, i P)$, $P\gg m_p$, is equivalent to a light-front formulation of the underlying theory \cite{Brodsky:1997de}.  This justifies $x\approx x_B$.

\begin{figure}[t]
\centerline{%
\includegraphics[clip, width=0.30\textwidth]{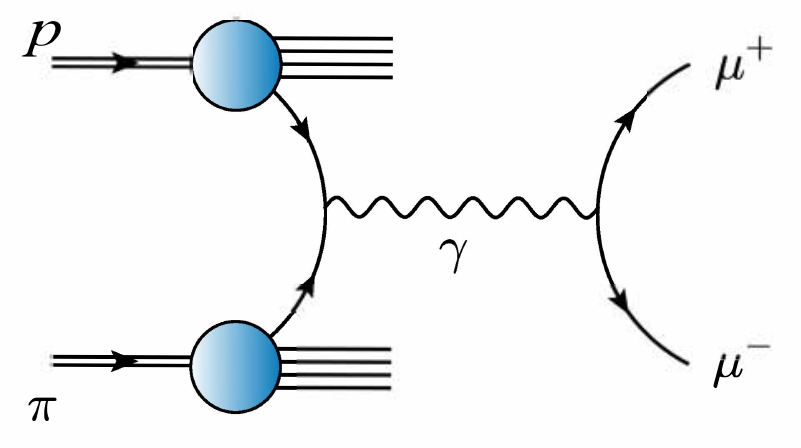}}
\caption{\label{FDY}
Pion + proton Drell-Yan process, \emph{e.g}., an antiquark in the pion annihilates with a quark in the proton to produce a muon + antimuon pair and a shower of hadrons -- $\pi p \to X + \mu^+ \mu^-$.  The invariant mass of the $\mu^+ \mu^-$ pair is $s \gg m_p^2$.
}
\end{figure}

The process-independence of DFs means that they can also be extracted from measurements of other high-energy processes.  For instance, pion parton DFs can be measured in pion + proton/nucleus ($\pi + p/A$) Drell-Yan experiments -- see Fig.\,\ref{FDY} and, \emph{e.g}., Ref.\,\cite{Ellis:1991qj}-[Ch.\,9].

Regarding such high-energy processes, QCD perturbation theory is used to express each associated cross-section as a convolution of (\emph{i}) elementary scattering processes involving the gluon and quark fields used to define the QCD Lagrangian density with (\emph{ii}) the in-hadron parton DFs.
The experimental energy scale at which the data is represented in terms of this convolution, $\zeta_E$, is also called the factorisation scale.  Asymptotic freedom in QCD \cite{Politzer:2005kc, Gross:2005kv, Wilczek:2005az} and the need to use perturbation theory to analyse the reaction means that $\zeta_E > m_p$ is required.

Supposing one has a valid treatment of (\emph{i}), then the parton DFs can be inferred via comparison between a measured cross-section and predictions of the convolution formula.  Of course, different measurements are often made at distinct energy scales and the inferred DFs depend on that scale; but within the context of perturbation theory, a DF obtained at scale $\zeta_{E^1}>m_p$ can be mapped into the DF measured at $\zeta_{E^2}>\zeta_{E^1}$ using perturbative scale evolution (DGLAP) equations \cite{Dokshitzer:1977sg, Gribov:1971zn, Lipatov:1974qm, Altarelli:1977zs}.

This fact somewhat obscures a critical issue.  The hard scattering reactions in the measurement processes can (and must) be treated using perturbation theory.  However, since DFs express properties of hadron wave functions, they are essentially nonperturbative in character.  So, whilst DFs may be inferred from data, they can only be predicted using a nonperturbative treatment of QCD.

For roughly forty years, models of hadron structure have been employed to estimate DFs \cite{Holt:2010vj}.  Such models are assumed to represent the target at some so-called hadron scale, $\zeta_{\cal H}< m_p$, and, from the beginning, the question has been \cite{Jaffe:1980ti}: At which value of $\zeta_{\cal H}$ should the result be viewed as valid?  This uncertainty means that, in almost all cases, $\zeta_{\cal H}$ is treated as a data fitting parameter; an optimal value being chosen so that, after DGLAP evolution, descriptive agreement with some selected data is achieved.  Consequently, both predictive power and the ability to use data to validate QCD are diminished.

An alternative to modelling can today be found in analyses of parton DF matrix elements using Euclidean space lattice-regularised QCD (lQCD) \cite{Lin:2017snn}, which are beginning to provide information about global and local properties of DFs.  In the lQCD scheme, the DF scale is known \emph{a priori}, being determined by the lattice setup.  However, challenges are posed by (\emph{i}) the needs to extrapolate to zero lattice spacing, infinite volume, and realistic quark masses, and (\emph{ii}) to map Euclidean-space lQCD results into the infinite momentum frame.  There is steady progress in these areas.

Herein, we describe an alternative to both modelling and lQCD.  The scheme exploits progress made with continuum Schwinger function methods (CSMs) in treating QCD \cite{Eichmann:2016yit, Burkert:2017djo, Fischer:2018sdj, Qin:2020rad}.  With CSMs, the challenge lies in a need to truncate the coupled set of quantum equations of motion when formulating the calculation of DF matrix elements.  It is being overcome by using symmetries, \emph{e.g}., Poincar\'e invariance and current conservation relations, to build sound approximations that deliver robust results \cite{Ding:2019lwe, Roberts:2021nhw, Binosi:2022djx, Papavassiliou:2022wrb, Ding:2022ows, Ferreira:2023fva}.

\medskip

\noindent\emph{2.$\;$Effective Charge and Hadron Scale}.
There are two principal elements in the formulation of all-orders evolution.  The first is the notion of an effective charge \cite{Grunberg:1980ja, Grunberg:1982fw, Deur:2023dzc}, \emph{viz}.\ a QCD running coupling, $\alpha_{\rm eff}(k^2)$, is defined by any chosen observable via the formula which expresses that observable to first-order in the perturbative coupling.  Thus defined, $\alpha_{\rm eff}(k^2)$ implicitly incorporates terms of arbitrarily high order in the perturbative  coupling.

Effective charges are typically process dependent, like $\alpha_{g_1}(k^2)$, defined via the Bjorken sum rule \cite{Deur:2022msf}.  On the other hand, any such effective charge has many valuable qualities, \emph{e.g}., it is: consistent with the QCD renormalisation group; renormalisation scheme independent; everywhere analytic and finite; and supplies an infrared completion of any standard running coupling.

Regarding DFs, other choices are possible and the following is proving efficacious: the effective charge $\alpha_{1\ell}(k^2)$ is that function which, when used to integrate the leading-order perturbative DGLAP equations, defines an evolution scheme for all parton DFs -- both unpolarised and polarised, and for any hadron -- that is all-orders exact.  This definition is broader than usual because it refers to an entire class of observables.  It is worth stressing that, although the pointwise form of $\alpha_{1\ell}(k^2)$ is largely irrelevant, the process-independent strong running coupling defined and computed in Refs.\,\cite{Binosi:2016nme, Cui:2019dwv} has all the required properties -- see, \emph{e.g}., Refs.\,\cite{Cui:2020dlm, Cui:2020tdf, Lu:2022cjx, Chang:2022jri}.

The second key to all-orders evolution is a specification of the hadron scale, $\zeta_{\cal H}<m_p$, \emph{i.e}., the point from which all-orders evolution should begin.  In order to eliminate all ambiguity, the natural choice is to associate $\zeta_{\cal H}$ with that scale at which all properties of a given hadron are carried by its valence quasiparticle d$^{\rm s}$of.  This means, \emph{e.g}., that all of a hadron's light-front momentum is carried by valence d$^{\rm s}$of at $\zeta_{\cal H}$; and, furthermore, that DFs associated with gluons and sea quarks are identically zero at $\zeta_{\cal H}$.

As an illustration, then, interpreted within the context of CSM treatments of QCD, this approach identifies $\zeta_{\cal H}$ as the scale at which strong interaction bound-state problems are most efficiently solved in terms of dressed-parton d$^{\rm s}$of.  Today, the characteristics of these dressed-gluons and -quarks are well understood \cite{Roberts:2021nhw, Binosi:2022djx, Papavassiliou:2022wrb, Ding:2022ows, Ferreira:2023fva}.

It is here worth recalling the approach to parton DFs introduced in Refs.\,\cite{Gluck:1995, Gluck:1998xa}-[GRV].  GRV supposed
that all DFs are nonzero at some low scale, $\mu_0 \approx \zeta_{\cal H}$, being a model parameter;
characterised those DFs by an array of parameters;
and used a perturbative QCD coupling to evolve the DFs to larger scales for comparison with results inferred from experiment.
In the GRV scheme, all parameters are varied so as to obtain a ``best fit'' to some body of data.
The point of contact between the GRV approach and all-orders evolution is the idea of generating all higher-scale DFs from some set of low (hadron) scale DFs; but that is where the similarity ends.
In the all-orders approach,
$\zeta_{\cal H}$ is not a parameter;
glue and sea DFs are identically zero at this scale;
the valence dof DFs are not parametrised -- instead, they are obtained from one or another nonperturbative calculation;
and specifying the pointwise form of $\alpha_{1\ell}(k^2)$ is typically unnecessary.

\medskip

\noindent\emph{3.$\;$Algebraic Evolution Equations: Unpolarised}.
Consider the unpolarised DF associated with a given dof, ${\mathpzc p}$, in hadron $H$: ${\mathpzc p}_H(x;\zeta_{\cal H})$.  Such DFs express an average over the polarisations/helicities of any given constituent dof.  The Mellin moments of this DF are defined as follows: 
\begin{equation}
\label{Mellin}
\langle x^n \rangle_{{\mathpzc p}_H}^{\zeta_{\cal H}} =
\int_0^1 dx\, x^n\,  {\mathpzc p}_H(x;\zeta_{\cal H})\,,\; n\in {\mathbb Z}^{\geq 0}\,.
\end{equation}

Given that any DF of physical interest is completely determined once all such Mellin moments are known, then it is sufficient to discuss all-orders evolution in terms of the behaviour of the moments in Eq.\,\eqref{Mellin}.  This enables one to arrive at an entirely algebraic formulation.  Further, there are two key constraints.  Conservation of baryon number in the Standard Model entails
\begin{equation}
\langle x^0 \rangle_{{\mathpzc p}_H}^{\zeta>\zeta_{\cal H}} =
\langle x^0 \rangle_{{\mathpzc p}_H}^{\zeta_{\cal H}} = n_{{\mathpzc p}\in H}\,,\;
\end{equation}
where $n_{{\mathpzc p}\in H}$ is the number of valence ${\mathpzc p}$ d$^{\rm s}$of in $H$, \emph{e.g}., $n_{{\mathpzc u}\in p}=2$ because there are two valence $u$ quarks in the proton; and, by definition, for any hadrons, $H$, $H^\prime$,
\begin{equation}
\label{MomConsevation}
\sum_{{\rm valence~d}^{\rm s}{\rm of}\, \in  H} \langle x \rangle_{{\mathpzc p}_H}^{\zeta_{\cal H}} = 1=
\sum_{{\rm valence~d}^{\rm s}{\rm of}\, \in  H^\prime} \langle x \rangle_{{\mathpzc p}_{H^\prime}}^{\zeta_{\cal H}}
\end{equation}
\emph{i.e}., in every hadron at $\zeta_{\cal H}$, all the light-front momentum is invested in valence d$^{\rm s}$of.

Written in terms of quark and antiquark DFs, the nonsinglet/valence and singlet DFs for a given quark flavour (${\mathpzc q} = u, d, s, c, \ldots)$ are, respectively,
\begin{subequations}
\begin{align}
{\cal V}_H^{\mathpzc q}(x;\zeta) & = {\mathpzc q}_H(x;\zeta) - \bar {\mathpzc q}_H(x;\zeta)\,, \\
\Sigma_H^{\mathpzc q}(x;\zeta) & = {\mathpzc q}_H(x;\zeta) + \bar {\mathpzc q}_H(x;\zeta)\,.
\end{align}
\end{subequations}
The associated sea distribution is ${\mathpzc S}_H^{\mathpzc q}(x;\zeta) = \Sigma_H^{\mathpzc q}(x;\zeta)-{\cal V}_H^{\mathpzc q}(x;\zeta) $ and we denote the glue distribution by ${\mathpzc g}_H(x;\zeta)$.

As reviewed elsewhere \cite{Ellis:1991qj}-[Ch.\,4], DGLAP evolution is defined by splitting functions.  Ignoring quark current-mass effects, there are four such functions: $P_{qq}(z)$, $P_{qg}(z)$, $P_{gq}(z)$, $P_{gg}(z)$, each of which expresses the probability for a given parton, ${\mathpzc p}$, to split into two other partons with light-front momentum fractions that are factors of $(z, 1-z)$ smaller than that of the parent.  In this context, for instance, $P_{qg}(z)$ is the probability that a gluon splits into a quark + antiquark pair, thereby transferring momentum from the glue DF into the singlet distribution of the given quark flavour.  The process becomes more effective as the number of active quark flavours increases.

Against this background and after some straightforward algebra, the all-orders evolution equations are:
{\allowdisplaybreaks
\begin{subequations}
\label{EqEvolution}
\begin{align}
\zeta^2 \frac{d}{d\zeta^2} \langle x^n \rangle_{{\mathpzc V}_H^{\mathpzc q}}^\zeta
& = - \frac{\alpha_{1\ell}(\zeta^2)}{4\pi}
\gamma_{qq}^n \langle x^n \rangle_{{\mathpzc V}_H^{\mathpzc q}}^\zeta \,,
\label{ValenceE}\\
\zeta^2 \frac{d}{d\zeta^2} \langle x^n \rangle_{\Sigma_H^{\mathpzc q}}^\zeta
& = - \frac{\alpha_{1\ell}(\zeta^2)}{4\pi}
\left\{
\gamma_{qq}^n \langle x^n \rangle_{\Sigma_H^{\mathpzc q}}^\zeta \right. \nonumber \\
& \quad \left.
+ 2 {\cal T}_{{\mathpzc q}g}^\zeta
\left[\gamma_{qg}^n
+ {\mathpzc B}_{{\mathpzc q}g}^{n\zeta}
\right]
\langle x^n \rangle_{{\mathpzc g}_H}^\zeta
\rule{0ex}{3ex}\right\}\,, \label{EqCalP}\\
\zeta^2 \frac{d}{d\zeta^2} \langle x^n \rangle_{{\mathpzc g}_H}^\zeta
& = - \frac{\alpha_{1\ell}(\zeta^2)}{4\pi}
\left\{ \gamma_{gq}^n \sum_{\mathpzc q}
 \langle x^n \rangle_{\Sigma_H^{\mathpzc q}}^\zeta \right. \nonumber \\
& \qquad
\left. + \gamma_{gg}^n \langle x^n \rangle_{{\mathpzc g}_H}^\zeta
\rule{0ex}{4ex}\right\}\,,  \label{EqCalPc}
\end{align}
\end{subequations}
where ${\mathpzc q}$ runs over all active quark flavours, $n_f=4$ $\Rightarrow$ $u$, $d$, $s$, $c$, and
$\gamma_{qq}^n$, $\gamma_{qg}^n$, $\gamma_{gq}^n$, $\gamma_{gg}^n$, $\beta_0=11-2 n_f/3$  are anomalous dimensions: 
\begin{subequations}
\label{eq:masslessADDGLAP}
\begin{align}
\gamma_{qq}^n &=\gamma_0^n \\
\label{eq:Dfn}
&= -\frac 4 3 \left[ 3 + \frac{2}{(n+1)(n+2)} - 4 \sum_{k=1}^{n+1} \frac 1 k  \right] \;, \\
\gamma_{qg}^n &= -\frac{4+n(3+n)}{(n+1)(n+2)(n+3)}\,,\\
%
\gamma_{gq}^n &= - \frac{8}{3} \frac{4+n(3+n)}{n (n+1) (n+2)}\,,\\
%
\gamma_{gg}^n &= -12 \left[
\frac{2(3+n(3+n))}{n(n+1)(n+2)(n+3)}
\right.  \left. 
- \sum_{k=1}^{n+1} \frac 1 k \right] - \beta_0\;.
\end{align}
\end{subequations}

These are textbook results for massless splitting functions, included here for ease of access.
Notably,
$\alpha_{1\ell}(\zeta^2)$ is positive definite and
$\gamma_{qq}^n$ increases monotonically from zero with $n$;
so, evolution works on ${\cal V}_H^{\mathpzc q}(x;\zeta)$ to shift support from large- to small-$x$ and into glue and sea DFs.

Equation~\eqref{EqCalP} contains two additional factors.
The first, ${\cal T}_{{\mathpzc q}g}^\zeta = \theta(\zeta - \epsilon_{\mathpzc q})$,
$\epsilon_{\mathpzc q}=\zeta_{\cal H} + \delta_{\mathpzc q}$,
is a quark flavour threshold function, which introduces the feature that a given quark flavour only participates in evolution once the resolving scale exceeds a value related to its effective mass.  Such a factor was exploited elsewhere \cite{Lu:2022cjx}, with $\delta_{{\mathpzc u},{\mathpzc d}}=0$, $\delta_{\mathpzc s}=0.1\,$GeV, $\delta_{\mathpzc c} = 0.9\,$GeV, to explain the small, yet nonzero $c+\bar c$ contribution to the proton light-front momentum fraction decomposition.

The second flavour-dependent factor in Eq.\,\eqref{EqCalP} derives from the following modification of the gluon splitting function:
\begin{equation}
\label{gluonsplit}
P_{qg}(z) \to P_{qg}(z;\zeta) = P_{qg}(z) +
\tfrac{\surd{3}}{2} C_1^{(1)}(1-2z) {\mathpzc P}_{\mathpzc q}(\zeta)\,,
\end{equation}
where $C_1^{(1)}$ is a Gegenbauer polynomial and
${\mathpzc P}_{\mathpzc q}(\zeta)= {\mathpzc b}_{{\mathpzc q} }/[1+(\zeta/\zeta_H-1)^2]$, with ${\mathpzc b}_{{\mathpzc q} }$ a number.
The correction term enables one to represent Pauli blocking in gluon splitting, \emph{viz}.\ following ideas in Ref.\,\cite{Field:1976ve}, gluon splitting into quark + antiquark pairs should be sensitive to the hadron's valence quark content so, \emph{e.g}.,  $g\to u+\bar u$ in the proton should be disfavoured relative to $g\to d+\bar d$.  The factor vanishes with increasing $\zeta$, expressing the diminishing influence of valence quarks as the proton's glue and sea content increases.  Using Eq.\,\eqref{gluonsplit}, one finds
\begin{equation}
\label{PBfunction}
{\mathpzc B}_{{\mathpzc q}g}^{n\zeta} = \frac{\surd 3\, n }{2+3 n + n^2}
{\mathpzc P}_{\mathpzc q}(\zeta)\,.
\end{equation}
${\mathpzc B}_{{\mathpzc q}g}^{0\zeta} =0$; hence, leaves $\langle x^0 \rangle$ unchanged, \emph{i.e}., Pauli blocking has no impact on baryon number.  Further, momentum conservation entails $\sum_{\mathpzc q}{\cal T}_{{\mathpzc q}g}^\zeta {\mathpzc b}_{{\mathpzc q}} = 0$.

Equation~\eqref{gluonsplit} was used elsewhere \cite{Chang:2022jri, Lu:2022cjx}, with
${\mathpzc b}_{{\mathpzc d} }=0.34=-{\mathpzc b}_{{\mathpzc u} }$, ${\mathpzc b}_{{\mathpzc s} } = 0 ={\mathpzc b}_{{\mathpzc c} }$, to explain the asymmetry of antimatter in the proton \cite{SeaQuest:2021zxb}-[SeaQuest].  In this case, the effect of Eq.\,\eqref{gluonsplit} was to shift momentum into $d+\bar d$ from $u+\bar u$, leaving the sum unchanged.

Equation~\eqref{ValenceE} is not explicitly coupled to Eqs.\,\eqref{EqCalP}, \eqref{EqCalPc}; so, can be solved in isolation: $\forall \zeta > \zeta_{\cal H}$,
\begin{subequations}
\label{ValenceE1}
\begin{align}
\langle x^n \rangle_{{\mathpzc V}_H^{\mathpzc q}}^\zeta
& = \langle x^n \rangle_{{\mathpzc V}_H^{\mathpzc q}}^{\zeta_{\cal H}}
[{\mathpzc E}_{\mathpzc V}(\zeta,\zeta_{\cal H})]^{\gamma_{qq}^n/\gamma_{0}^1} \,,\\
{\mathpzc E}_{{\mathpzc V}}(\zeta,\zeta_{\cal H}) & =
\exp\left[\frac{\gamma_{0}^1}{4\pi} \int^{\ln\zeta_{\cal H}^2}_{\ln \zeta^2}\!  dt\,
\alpha_{1\ell}({\rm e}^t)
\right]\,,
\end{align}
\end{subequations}
where we have exploited universality of $\alpha_{1\ell}$.
\emph{N.B}.\ With increasing $\zeta$, ${\mathpzc E}_{{\mathpzc V}}(\zeta,\zeta_{\cal H})$ falls logarithmically (with exponent $\gamma_0^1/\beta_0$) on $\zeta\gg\zeta_{\cal H}$.

Regarding the solutions for glue and sea, one proceeds by
defining moments of the net singlet distribution
\begin{equation}
\langle x^n \rangle_{\Sigma_{H}}^\zeta = \sum_{\mathpzc q} \langle x^n \rangle_{\Sigma_{H}^{\mathpzc q}}^\zeta \,,
\end{equation}
in terms of which, using momentum conservation, the solutions of Eqs.\,\eqref{EqCalP}, \eqref{EqCalPc} can be written:
{\allowdisplaybreaks
\begin{subequations}
\label{Solutions}
\begin{align}
& \left[
\begin{array}{l}
\langle x^n \rangle_{\Sigma_{H}}^\zeta \\
\langle x^n \rangle_{{\mathpzc g}_{H}}^\zeta
\end{array}
\right]
=
\left\{
\begin{array}{c}
{\mathpzc T}_2 \\
{\mathpzc T}_3 \\
{\mathpzc T}_4 \\
\end{array}
\right\}
\left[
\begin{array}{l}
\langle x^n \rangle_{\Sigma_{H}}^{\zeta_{\cal H}} \\
\langle x^n \rangle_{{\mathpzc g}_{H}}^{\zeta_{\cal H}}
\end{array}
\right]\,,  \\
&
\left\{
\begin{array}{c}
{\mathpzc T}_2 \\
{\mathpzc T}_3 \\
{\mathpzc T}_4 \\
\end{array}
\right\}
=
\left\{
\begin{array}{c}
\left. {\mathpzc E}_{\mathpzc S}^2(\zeta,\zeta_{\cal H})\right|_{\zeta<\epsilon_{\mathpzc s}}\\
\left.
{\mathpzc E}_{\mathpzc S}^3(\zeta,\epsilon_{\mathpzc s})
{\mathpzc E}_{\mathpzc S}^2(\epsilon_{\mathpzc s},\zeta_{\cal H})
\right|_{\epsilon_{\mathpzc s} < \zeta<\epsilon_{\mathpzc c}}  \\
\left.
{\mathpzc E}_{\mathpzc S}^4(\zeta,\epsilon_{\mathpzc c})
{\mathpzc E}_{\mathpzc S}^3(\epsilon_{\mathpzc c},\epsilon_{\mathpzc s})
{\mathpzc E}_{\mathpzc S}^2(\epsilon_{\mathpzc s},\zeta_{\cal H})
\right|_{\epsilon_{\mathpzc c} < \zeta}
\end{array}
\right\}\,.
\end{align}
\end{subequations}
Here
\begin{equation}
{\mathpzc E}^n_{\mathpzc S}(\zeta_2,\zeta_1) =
\left[
\begin{array}{cc}
\alpha_+^n {\mathpzc E}^n_- + \alpha_-^n {\mathpzc E}^n_+
    & \beta^n_{g \Sigma}[{\mathpzc E}^n_- - {\mathpzc E}^n_+] \\
\beta^n_{\Sigma g}[{\mathpzc E}^n_- - {\mathpzc E}^n_+]
    & \alpha_-^n {\mathpzc E}^n_- + \alpha_+^n {\mathpzc E}^n_+
\end{array}
\right]\,, \label{ESmatrix}
\end{equation}
where
\begin{subequations}
\begin{align}
\alpha_\pm^n & = \pm \frac{\lambda_\pm^n - \gamma_{qq}^n}{\lambda_+^n-\lambda_-^n}\,,\;
{\mathpzc E}^n_\pm  = [{\mathpzc E}_{\mathpzc V}(\zeta_2,\zeta_1)]^{\lambda_\pm^n/\gamma_0^1} \,, \\
\beta_{\Sigma g} & = - \frac{2 n_f \gamma_{qg}^n}{\lambda_+^n-\lambda_-^n}\,,\;
\beta_{g\Sigma}  =
\frac{(\lambda_+^n - \gamma_{qq}^n)(\lambda_-^n - \gamma_{qq}^n)}
{2 n_f \gamma_{qg}^n (\lambda_+^n-\lambda_-^n)}\,,
\end{align}
\end{subequations}
and
$\lambda_\pm^n = \tfrac{1}{2} {\rm tr} \Gamma^n
\pm \tfrac{1}{2}\sqrt{[{\rm tr} \Gamma^n]^2 - 4 \det \Gamma^n}$
are eigenvalues of the matrix of anomalous dimensions:
\begin{equation}
\Gamma^n = \left[
\begin{array}{cc}
\gamma_{qq}^n & 2 n_f \gamma_{qg}^n \\
\gamma_{gq}^n & \gamma_{gg}
\end{array}
\right]\,.
\label{GammaUnP}
\end{equation}
\emph{N.B}.\ $\alpha_+^n+\alpha_-^n =1$, $\alpha_+^n \alpha_-^n = \beta_{g\Sigma}^n\beta_{\Sigma g}^n$; so, ${\mathpzc E}^n_{\mathpzc S}(\zeta,\zeta) = \mathbb I$.
}

It is now plain that, within the all-orders scheme, the evolution kernel for all DFs is that which governs evolution of the valence DFs.  Using Eq.\,\eqref{ValenceE1}, one obtains the characteristic equations of all-orders evolution:
\begin{subequations}
\label{MasterEq}
\begin{align}
{\mathpzc E}_{{\mathpzc V}}(\zeta_2,\zeta_1) & =
\frac{\langle x \rangle_{{\mathpzc V}_p^{\mathpzc u}}^{\zeta_2}
+ \langle x \rangle_{{\mathpzc V}_p^{\mathpzc d}}^{\zeta_2}}
{\langle x \rangle_{{\mathpzc V}_p^{\mathpzc u}}^{\zeta_1}
+ \langle x \rangle_{{\mathpzc V}_p^{\mathpzc d}}^{\zeta_1}}\,,
\label{MasterEqI}\\
{\mathpzc E}_{{\mathpzc V}}(\zeta,\zeta_{\cal H}) & =
\langle x \rangle_{{\mathpzc V}_p^{\mathpzc u}}^{\zeta}
+ \langle x \rangle_{{\mathpzc V}_p^{\mathpzc d}}^{\zeta}=: \langle x \rangle_{{\mathpzc V}_p}^\zeta \,,
\end{align}
\end{subequations}
where the last line follows from Eq.\,\eqref{MomConsevation}.
These identities are true for any hadron (meson or baryon): the right-hand side is just the ratio of the sums of the light-front momentum fractions of the valence d$^{\rm s}$of at two different scales.
Evidently:
(\emph{i}) the all-orders evolution kernel is independent of the form of $\alpha_{1\ell}(k^2)$;
and
(\emph{ii}) given any DF at one scale, $\zeta_1$, then its form at any other scale, $\zeta_2$, is completely determined once the valence dof momentum fraction at $\zeta_2$ is known.
Clearly, (\emph{ii}) entails that all DFs are determined by the valence dof DFs at $\zeta_{\cal H}$.
This is given weight by Eq.\,\eqref{MomConsevation}, which means that all evolution can be referred back to that single hadron whose properties are most readily measurable, \emph{e.g}., the proton.

It is interesting to temporarily suppress quark flavour thresholds, \emph{i.e}., in Eq.\,\eqref{EqEvolution}, $\forall {\mathpzc q}$, write
${\cal T}_{{\mathpzc q}g}^\zeta = \theta(\zeta-\zeta_{\cal H})$.  Then, exploiting the character of the hadron scale, \emph{viz}.\ $\langle x^n \rangle_{{\mathpzc g}_{H}}^{\zeta_{\cal H}} \equiv 0$, Eq.\,\eqref{Solutions} becomes:
\begin{equation}
\label{uvlimit4}
\left[
\begin{array}{l}
\langle x^n \rangle_{\Sigma_{H}}^\zeta \\
\langle x^n \rangle_{{\mathpzc g}_{H}}^\zeta
\end{array}
\right]
=
\left[
\begin{array}{c}
\alpha_+^n {\mathpzc E}^n_- + \alpha_-^n {\mathpzc E}^n_+ \\
\beta^n_{\Sigma g}[{\mathpzc E}^n_- - {\mathpzc E}^n_+]
\end{array}
\right] \langle x^n \rangle_{\Sigma_{H}}^{\zeta_{\cal H}}\,.
\end{equation}

Specialising to $n=1$: $\lambda_+^1=56/9$, $\lambda_-^1 = 0$, $\alpha_+^1=\beta_{\Sigma g}^1=3/7$,
$\alpha_-^1=\beta_{g\Sigma }^1=4/7$; then using $\langle x \rangle_{\Sigma_{H}}^{\zeta_{\cal H}}=1$ and Eqs.\,\eqref{MasterEq}, one arrives at the following:
\begin{equation}
\label{UVfractions}
\begin{array}{ll}
\langle x \rangle_{\Sigma_{H}}^\zeta & = \tfrac{3}{7}
+ \tfrac{4}{7}[ \langle x \rangle_{{\mathpzc V}_p}^\zeta]^{7/4}, \\[1ex]
\langle x \rangle_{{\mathpzc g}_{H}}^\zeta
& = \tfrac{4}{7} (1 - [\langle x \rangle_{{\mathpzc V}_p}^\zeta]^{7/4} )\,,
\end{array}
\end{equation}
from which one recovers the textbook $\zeta \gg\zeta_{\cal H}$ results for quark and gluon momentum fractions in any hadron.

Suppose now that Eqs.\,\eqref{uvlimit4}, \eqref{UVfractions} are valid $\forall \zeta > \zeta_{\cal H}$ and reinterpret them as statements about the DFs $x \Sigma_{H}(x;\zeta)$, $x {\mathpzc g}_{H}(x;\zeta)$.  Then, in the limit $\zeta/\zeta_{\cal H}\to \infty$, the zeroth moments of these distributions are nonzero constants ($3/7,4/7$) and all other moments vanish; hence,
\begin{equation}
x \Sigma_{H}(x;\zeta) \stackrel{\zeta_{\cal H}/\zeta \simeq 0}{\approx} \tfrac{3}{7} \delta(x)\,, \;
x g_{H}(x;\zeta) \stackrel{\zeta_{\cal H}/\zeta \simeq 0}{\approx} \tfrac{4}{7} \delta(x)\,.
\end{equation}

Restoring threshold factors, one readily obtains the $\zeta > \epsilon_{\mathpzc c}$ generalisation of Eqs.\,\eqref{UVfractions}:
\begin{equation}
\label{UVfractionsII}
\begin{array}{ll}
\langle x \rangle_{\Sigma_{H}}^\zeta & = \tfrac{3}{7}
+ \tau(\epsilon_{\mathpzc c},\epsilon_{\mathpzc s}) [ \langle x \rangle_{{\mathpzc V}_p}^\zeta]^{7/4}, \\
\langle x \rangle_{{\mathpzc g}_{H}}^\zeta
& = \tfrac{4}{7} - \tau(\epsilon_{\mathpzc c},\epsilon_{\mathpzc s})[\langle x \rangle_{{\mathpzc V}_p}^\zeta]^{7/4} \,,
\end{array}
\end{equation}
where
\begin{align}
& \tau(\epsilon_{\mathpzc c},\epsilon_{\mathpzc s})  =
-\tfrac{12}{175}  [ \langle x \rangle_{{\mathpzc V}_p}^{\epsilon_{\mathpzc c}}]^{-\tfrac{7}{4}} \nonumber \\
& -\tfrac{24}{275} [ \langle x \rangle_{{\mathpzc V}_p}^{\epsilon_{\mathpzc c}}]^{-\tfrac{3}{16}}
 [ \langle x \rangle_{{\mathpzc V}_p}^{\epsilon_{\mathpzc s}}]^{-\tfrac{25}{16}}
+ \tfrac{8}{11}  [ \langle x \rangle_{{\mathpzc V}_p}^{\epsilon_{\mathpzc c}}
\langle x \rangle_{{\mathpzc V}_p}^{\epsilon_{\mathpzc s}}]^{-\tfrac{3}{16}}.
\label{Evolutiontau}
\end{align}
Formulae valid between thresholds may be obtained progressively by setting $\epsilon_{\mathpzc c}\to \zeta$ and then $\epsilon_{\mathpzc s}\to \zeta$; and Eqs.\,\eqref{UVfractions} are recovered by setting $\epsilon_{{\mathpzc c},\mathpzc s}\to \zeta_{\cal H}$.
\emph{N.B}.\ Equations~\eqref{UVfractionsII}, \eqref{Evolutiontau} highlight that the presence of thresholds introduces some sensitivity to the pointwise behaviour of $\alpha_{1\ell}(k^2)$ because the momentum fraction at $\zeta$ depends on its value at distinct lower scales greater than $\zeta_{\cal H}$.

Return now to Eqs.\,\eqref{EqEvolution} and consider the sea quark DF for any given flavour, then
\begin{equation}
\langle x^n \rangle_{{\mathpzc S}_H^{\mathpzc q}}^\zeta =
\langle x^n \rangle_{{\Sigma}_H^{\mathpzc q}}^\zeta
- \langle x^n \rangle_{{\mathpzc V}_H^{\mathpzc q}}^\zeta\,.
\end{equation}
Using this identity, the difference Eq.\,\eqref{EqCalP}\,$-$\,Eq.\,\eqref{ValenceE} yields the following
\begin{align}
\left[\zeta^2 \frac{d}{d\zeta^2} \right. & \left. + \frac{\alpha_{1\ell}(\zeta^2)}{4\pi} \gamma_{qq}^n \right]
\langle x^n \rangle_{{\mathpzc S}_H^{\mathpzc q}}^\zeta \nonumber \\
& = - \frac{\alpha_{1\ell}(\zeta^2)}{4\pi}
2 {\cal T}_{{\mathpzc q}g}^\zeta
\left[\gamma_{qg}^n
+ {\mathpzc B}_{{\mathpzc q}g}^{n\zeta}
\right]
\langle x^n \rangle_{{\mathpzc g}_H}^\zeta\,.  \label{DFSea}
\end{align}
Evidently, the sea quark DFs are determined once the glue DF is known and the latter is solved independently in terms of the hadron's valence dof moment.  Further algebra reveals the solution of Eq.\,\eqref{DFSea} ($z= {\rm e}^{t/2}$):
\begin{subequations}
\label{SeaBlocking}
\begin{align}
\langle x^n \rangle_{{\mathpzc S}_H^{\mathpzc q}}^\zeta
& = - {\cal T}_{{\mathpzc q}g}^\zeta
\frac{\gamma_{qg}^n }{2\pi} \int_{\ln \epsilon_{\mathpzc q}^2}^{\ln \zeta^2} \!
dt\,  \bigg[\alpha_{1\ell}(z^2)
\langle x^n \rangle_{{\mathpzc g}_H}^{z} \nonumber \\
& \qquad\times
[\langle x \rangle_{{\mathpzc V}_H}^\zeta/\langle x \rangle_{{\mathpzc V}_H}^z]^{\gamma_{qq}^n/\gamma_{0}^1} \bigg]
 -  \tilde {\cal T}_{{\mathpzc q}g}^\zeta  {\mathbb B}^{n\zeta}_{\mathpzc q} \,, \\
{\mathbb B}^{n\zeta}_{\mathpzc q} & = -
\frac{1}{2\pi} \int_{\ln \epsilon_{\mathpzc q}^2}^{\ln \zeta^2} \!
dt\,  \alpha_{1\ell}(z^2) \nonumber \\
& \qquad \times
 {\mathpzc B}_{{\mathpzc q}g}^{n z}
 \langle x^n \rangle_{{\mathpzc g}_H}^{z}
[\langle x \rangle_{{\mathpzc V}_H}^\zeta/\langle x \rangle_{{\mathpzc V}_H}^z]^{\gamma_{qq}^n/\gamma_{0}^1} .
\end{align}
\end{subequations}
If present, the Pauli blocking factor introduces some additional sensitivity to the pointwise form of $\alpha_{1\ell}(k^2)$.

It is again worth considering the special case of $n=1$, \emph{i.e}., the light-front fraction of a given hadron's momentum stored in each flavour component of the sea.  Using the above identities, straightforward algebra confirms momentum conservation:
{\allowdisplaybreaks
\begin{subequations}
\begin{align}
\langle x\rangle_{{\mathpzc S}_H}^\zeta
= \sum_{\mathpzc q}\langle x\rangle_{{\mathpzc S}_H^{\mathpzc q}}^\zeta
& = 1 - \langle x \rangle_{{\mathpzc g}_H}^{\zeta} - \sum_{\mathpzc q} \langle x \rangle_{{\mathpzc V}^q_H}^\zeta \\
& 
= 1 - \langle x \rangle_{{\mathpzc g}_H}^{\zeta} - \langle x \rangle_{{\mathpzc V}_H}^\zeta \,,
\end{align}
\end{subequations}
From here, additional algebra delivers the following $\zeta > \epsilon_{\mathpzc c}$ predictions
($\lambda_{{\mathpzc g}_H}^{\zeta}=1 - \langle x \rangle_{{\mathpzc g}_H}^{\zeta})$:
\begin{subequations}
\label{SeaFractions}
\begin{align}
\langle x\rangle_{{\mathpzc S}_H^{{\mathpzc u},{\mathpzc d}}}^\zeta
& = \tfrac{1}{4}\lambda_{{\mathpzc g}_H}^{\zeta}
+ \tfrac{1}{12}\lambda_{{\mathpzc g}_H}^{\epsilon_{\mathpzc c}}{\mathpzc E}_{\mathpzc V}(\zeta,\epsilon_{\mathpzc c}) \nonumber \\
& \quad + \tfrac{1}{6} \lambda_{{\mathpzc g}_H}^{\epsilon_{\mathpzc s}}{\mathpzc E}_{\mathpzc V}(\zeta,\epsilon_{\mathpzc s})
-\tfrac{1}{2} \langle x \rangle_{{\mathpzc V}_H}^\zeta
- {\mathbb B}^{1\zeta}_{{\mathpzc u},{\mathpzc d}}\,,
\\
\langle x\rangle_{{\mathpzc S}_H^{\mathpzc s}}^\zeta
& = \tfrac{1}{4}\lambda_{{\mathpzc g}_H}^{\zeta}
+ \tfrac{1}{12}\lambda_{{\mathpzc g}_H}^{\epsilon_{\mathpzc c}}{\mathpzc E}_{\mathpzc V}(\zeta,\epsilon_{\mathpzc c}) \nonumber \\
& \quad
- \tfrac{1}{3}\lambda_{{\mathpzc g}_H}^{\epsilon_{\mathpzc s}}{\mathpzc E}_{\mathpzc V}(\zeta,\epsilon_{\mathpzc s})
- {\mathbb B}^{1\zeta}_{\mathpzc s}
\,, \\
\langle x\rangle_{{\mathpzc S}_H^{\mathpzc c}}^\zeta
& = \tfrac{1}{4}\lambda_{{\mathpzc g}_H}^{\zeta}
- \tfrac{1}{4}\lambda_{{\mathpzc g}_H}^{\epsilon_{\mathpzc c}}{\mathpzc E}_{\mathpzc V}(\zeta,\epsilon_{\mathpzc c})
- {\mathbb B}^{1\zeta}_{\mathpzc c} \,,
\end{align}
\end{subequations}
with $\langle x \rangle_{{\mathpzc g}_H}^{\zeta}$ given in Eq.\,\eqref{UVfractionsII}.
Again, formulae valid between thresholds may be obtained progressively by setting $\epsilon_{\mathpzc c}\to \zeta$ and then $\epsilon_{\mathpzc s}\to \zeta$.
}

It is worth illustrating the impact of such Pauli blocking using the pion as an exemplar.  Below the $s$ quark threshold, gluon splitting produces $u+\bar u$ and $d+\bar d$ with some probability and only these combinations.  On the other hand, above the threshold, $s+\bar s$ production is both possible and preferred.  This may be expressed by writing
$\tilde{\cal T}_{{\mathpzc u}g}^\zeta = \tilde{\cal T}_{{\mathpzc d}g}^\zeta = \tilde{\cal T}_{{\mathpzc s}g}^\zeta = \theta(\zeta-\epsilon_{\mathpzc s})$ in Eqs.\,\eqref{SeaBlocking} and ${\mathpzc b}_{\mathpzc s} = \overline{\mathpzc b}= -2 {\mathpzc b}_{\mathpzc u}=-2 {\mathpzc b}_{\mathpzc d}$ in Eq.\,\eqref{gluonsplit}.
Then, using the process-independent charge described in Ref.\,\cite{Cui:2020dlm}-[Sec.\,3] for $\alpha_{1\ell}(k^2)$, associated parton DF predictions in Ref.\,\cite{Lu:2022cjx}, and the value ${\mathpzc b}_{\mathpzc u}=  0.34(9)$ required to explain the SeaQuest data \cite{Chang:2022jri}, one finds
\begin{equation}
\label{sseaPB}
\langle x\rangle_{{\mathpzc S}_\pi^{\mathpzc s}}^{\epsilon_{\mathpzc c}}
= 1.29(11) \langle x\rangle_{{\mathpzc S}_\pi^{\mathpzc u}}^{\epsilon_{\mathpzc c}}
= 1.29(11) \langle x\rangle_{{\mathpzc S}_\pi^{\mathpzc d}}^{\epsilon_{\mathpzc c}}\,.
\end{equation}
(Ignoring flavour thresholds and Pauli blocking, the right-hand side would be $1$.)
Thus, the light-front momentum fraction carried by $s$ sea quarks at $\zeta=\epsilon_{\mathpzc c}$ exceeds that carried by each flavour in the light quark sea; and, at the $c$ quark threshold, Pauli blocking has shifted 12(3)\% of the momentum from each flavour in the light quark sea into the $s$ quark sea.
These outcomes mean that the same level of Pauli blocking which explains the asymmetry of antimatter in the proton leads to a marked excess of $s+\bar s$ relative to $u+\bar u$ and $d+\bar d$ in the pion sea.
Qualitatively similar outcomes relative to their valence quark/dof content should be expected in other hadrons.

\medskip

\noindent\emph{4.$\;$Algebraic Evolution Equations: Polarised}.
The polarised DF $\Delta {\mathpzc p}(x;\zeta)$ expresses the light-front helicity distribution of a parton $\mathpzc p$ carrying a light-front fraction $x$ of the hadron's momentum at the scale $\zeta$.  Such DFs measure the difference between the light-front number density of partons with helicity parallel to that of the hadron and those with antiparallel helicity.  Thus, polarised DFs describe the contribution of a given parton species to the hadron's total spin, $J$.

Extraction of polarised DFs requires DIS(-like) measurements with longitudinally polarised beams and targets.  These demands have limited the available empirical information.  Nevertheless, results from measurements in Ref.\,\cite{EuropeanMuon:1987isl} ignited the ``proton spin crisis'' with a finding which, interpreted naively, suggests that quarks carry little of the proton spin -- see, \emph{e.g}., Ref.\,\cite{Anselmino:1994gn, Deur:2018roz}.

Concerning all-orders evolution, Eqs.\,\eqref{ValenceE1}\,-\,\eqref{ESmatrix} apply equally to polarised DFs, so long as the anomalous dimensions are amended appropriately.
Regarding Ref.\,\cite{Altarelli:1977zs} and comparing Eqs.\,(71)-(74) with Eqs.\,(100), one sees that the nonsinglet anomalous dimension is unchanged.  Hence, evolution of $\langle x^n \rangle_{\Delta{\mathpzc V}_H^{\mathpzc q}}^\zeta$ is also given by Eqs.\,\eqref{ValenceE1}.
The other three entries in Eq.\,\eqref{GammaUnP} change: $\gamma \to \tilde\gamma$, with
\begin{subequations}
\label{eq:masslessADDGLAP:Pol}
\begin{align}
\tilde \gamma_{qg}^n &= - 2 n_f \frac{n}{(n+1)(n+2)}\,, \\
\tilde \gamma_{gq}^n &= - \frac{8}{3} \frac{n+3}{(n+1)(n+2)}\,, \\
\tilde \gamma_{gg}^n &= 6 \left[\frac{2n}{(n+1)(n+2)} + 2 \sum_{k=1}^n\frac{1}{k}\right]
-\beta_0\,.
\end{align}
\end{subequations}
(The sum in the last line is zero for $n=0$.)
Thereafter, all steps are qualitatively identical to the unpolarised case.

Regarding polarised DFs, $n=0$ is of principal interest because it relates to the proton spin.  In this case, since $\gamma_{qq}^0=0$, then, under all-orders evolution:
\begin{equation}
\forall \zeta > \zeta_{\cal H}, \;
\langle x^0 \rangle_{\Delta{\mathpzc V}_H^{\mathpzc q}}^\zeta
= \langle x^0 \rangle_{\Delta{\mathpzc V}_H^{\mathpzc q}}^{\zeta_{\cal H}}\,.
\end{equation}
Moreover, $\tilde\gamma_{qq}^0=\gamma_{qq}^0 =0 $,
$\tilde\gamma_{qg}^0=0$; hence, neglecting flavour thresholds,
\begin{subequations}
\label{uvlimit4Pol}
\begin{align}
a_{0H}^\zeta := \langle x^0 \rangle_{\Delta\Sigma_{H}}^\zeta & =
\langle x^0 \rangle_{\Delta\Sigma_{H}}^{\zeta_{\cal H}}\,, \label{uvlimit4PolA} \\
\Delta G_{H}^{\zeta} :=\langle x^0 \rangle_{\Delta{\mathpzc g}_{H}}^\zeta & =
a_0^\zeta \,
\tfrac{12}{25}
\left[[ \langle x \rangle_{{\mathpzc V}_H}^\zeta ]^{-75/32} - 1\right]\,. \label{uvlimit4PolB}
\end{align}
\end{subequations}
These novel identities state (\emph{i}) that the valence and singlet quark polarised DFs are scale invariant and (\emph{ii}), remarkably, the polarised gluon DF grows as the inverse of the net unpolarised valence parton DF with a constant of proportionality fixed by the singlet quark polarised DF.

It is worth illustrating the content of Eqs.\,\eqref{uvlimit4Pol} using the proton.
Regarding the polarised singlet quark DF, if one assumes SU$(3)$-flavour symmetry in analyses of the axial charges of octet baryons, then these charges are expressed in terms of two low-energy constants \cite{Cabibbo:2003cu}-[Table~1], $D$, $F$; and
$\langle x^0 \rangle_{\Delta\Sigma_{p}^{\mathpzc u}}^{\zeta_{\cal H}} = 2 F$,
$\langle x^0 \rangle_{\Delta\Sigma_{p}^{\mathpzc d}}^{\zeta_{\cal H}} = F-D$.
Using empirical information \cite{Workman:2022ynf, ChenChen:2022qpy}, one finds
$D= 0.774(26)$, $F=0.503(27)$;
hence,
$\langle x^0 \rangle_{\Delta\Sigma_{p}^{\mathpzc u}}^{\zeta_{\cal H}} = 1.01(5)$,
$\langle x^0 \rangle_{\Delta\Sigma_{p}^{\mathpzc d}}^{\zeta_{\cal H}} = -0.27(5)$,
and
\begin{equation}
\label{Empircalxo0}
a_{0p}^\zeta = \langle x^0 \rangle_{\Delta\Sigma_{p}}^{\zeta} = 0.74(11) \quad \mbox{\rm empirical} \,.
\end{equation}

Using Eq.\,\eqref{uvlimit4PolB}, a result for the integrated strength of the polarised gluon DF follows immediately once a value of $\langle x \rangle_{{\mathpzc V}_p}^\zeta $ is in hand.  One might choose to use a value obtained from one or another of the existing phenomenological global fits to high-energy data.  A typical result \cite{Hou:2019efy}-[CT18] is listed in Position (2,A) of Table~\ref{tablepolarised}.  Such fits may underestimate the valence quark momentum fraction, lodging too much momentum with the sea \cite{Cui:2021mom}.  Notwithstanding that, combined with Eq.\,\eqref{Empircalxo0} via Eq.\,\eqref{uvlimit4PolB}, this fraction yields the polarised gluon moment in Position (6,A) of Table~\ref{tablepolarised}: it is large because the valence quark momentum fraction is small.

\begin{table}[t]
\caption{\label{tablepolarised}
Block 1 -- Rows~1 and 2: Source and value of the valence quark moment.
Block 2 -- Rows~3 and 4: Source and value of the polarised singlet quark moment.
Block 3 -- Row~5 and 6: equation used to compute the listed value of the polarised gluon moment.
Block 4 -- Row 7: net singlet moment after non-Abelian anomaly correction, Eq.\,\eqref{Eqa0E}.  This value can be compared with that reported in Ref.\,\cite{COMPASS:2016jwv}-[COMPASS]: $0.32(07)$.
Where relevant, the Mellin moments are given at the scale $\zeta=\sqrt{3}\,$GeV$\,=:\zeta_{\rm C}$.
 }
\begin{center}
\begin{tabular*}
{\hsize}
{
l@{\extracolsep{0ptplus1fil}}
c@{\extracolsep{0ptplus1fil}}
c@{\extracolsep{0ptplus1fil}}
c@{\extracolsep{0ptplus1fil}}
c@{\extracolsep{0ptplus1fil}}}\hline\hline
& A & B & C & D \\\hline
& CT18 & Ref.\,\cite{Lu:2022cjx} & Ref.\,\cite{Lu:2022cjx} & Ref.\,\cite{Lu:2022cjx} \\
$\rule{0ex}{3ex} \langle x \rangle_{{\mathpzc V}_p}^\zeta $ & $0.42(01)$ & $0.49(02)$ & $0.49(02)$ &$0.49(02)$ \\[1.1ex]\hline\hline
& Eq.\,\eqref{Empircalxo0} & Eq.\,\eqref{Empircalxo0} & Ref.\,\cite{ChenChen:2022qpy} & Ref.\,\cite{ChenChen:2022qpy} \\
$ a_{0p}^\zeta$
& $0.74(11)$ &$0.74(11)$ & $0.65(02)$ & $0.65(02)$ \\[1.1ex]\hline\hline
& Eq.\,\eqref{uvlimit4PolB} & Eq.\,\eqref{uvlimit4PolB} & Eq.\,\eqref{uvlimit4PolB} & Eq.\,\eqref{PolGlueThresh} \\
$\Delta G_p^\zeta$
& $2.27(30)$ &$1.50(25)$ & $1.33(15)$ &  $1.41(16)$ \\[1.1ex]\hline\hline
$\tilde a_{0p}^\zeta$ & $0.20(11)$ & $0.38(11)$ & $0.33(04)$ & $0.32(04)$ \\\hline\hline
\end{tabular*}
\end{center}
\end{table}

Experiments aimed at extracting the quark contribution to the proton spin measure the following non-Abelian anomaly corrected quantity \cite{Altarelli:1988nr}:
\begin{equation}
\label{Eqa0E}
\tilde a_{0p}^{\zeta} = a_{0p}^\zeta
- n_f \frac{\hat \alpha(\zeta)}{2\pi}\Delta G_p^\zeta \,.
\end{equation}
Combining the results in Table~\ref{tablepolarised}\,--\,Column~A, one arrives at the value in Position (7,A).

In contrast, one may use the CSM prediction for the proton valence quark momentum fraction in combination with either the empirical value of $a_{0p}^\zeta$, Eq.\,\eqref{Empircalxo0}, or the prediction from Ref.\,\cite{ChenChen:2022qpy}.  The results thereby obtained are listed in Columns~B and C of Table~\ref{tablepolarised}.

Reintroducing quark flavour thresholds, Eq.\,\eqref{uvlimit4PolA} is unchanged, whereas on $\zeta > \epsilon_{\mathpzc c}$, Eq.\,\eqref{uvlimit4PolB} becomes, referring to Eq.\,\eqref{MasterEq}:
\begin{align}
\Delta G_{H}^\zeta & = a_{0H}^\zeta
\left\{
\tfrac{12}{25} \left[[{\mathpzc E}_{{\mathpzc V}}(\zeta,\epsilon_{\mathpzc c})]^{-75/32}-1 \right]
\right.  \nonumber \\
& \quad + \tfrac{4}{9} [{\mathpzc E}_{{\mathpzc V}}(\zeta,\epsilon_{\mathpzc c})]^{-75/32}
\left[[{\mathpzc E}_{{\mathpzc V}}(\epsilon_{\mathpzc c},\epsilon_{\mathpzc s})]^{-81/32}-1 \right]
\nonumber \\
& \quad +
\tfrac{12}{29}
{\mathpzc E}_{{\mathpzc V}}(\zeta,\epsilon_{\mathpzc c})]^{-75/32}\,
[{\mathpzc E}_{{\mathpzc V}}(\epsilon_{\mathpzc c},\epsilon_{\mathpzc s})]^{-81/32} \nonumber \\
& \qquad \times
\left.
\left[[{\mathpzc E}_{{\mathpzc V}}(\epsilon_{\mathpzc s},\zeta_{\cal H})]^{-87/32}-1 \right]
\right\}\,.
\label{PolGlueThresh}
\end{align}
As usual, formulae valid between thresholds may be obtained progressively by setting $\epsilon_{\mathpzc c}\to \zeta$ and then $\epsilon_{\mathpzc s}\to \zeta$.  The no-threshold result is recovered by setting $\epsilon_{{\mathpzc c},{\mathpzc s}} \to \zeta_{\cal H}$.

Using Eq.\,\eqref{PolGlueThresh} and, for $\alpha_{1\ell}(k^2)$, the process-independent charge described in Ref.\,\cite{Cui:2020dlm}-[Sec.\,3] along with the entries in Table~\ref{tablepolarised}\,--\,Column~D, one obtains the results for $\Delta G_{p}^\zeta$ in Position (5,D).  They correspond to the CSM predictions in Ref.\,\cite{Cheng:2023kmt}, after accounting for extrapolation of
$\int_{x_0}^1 dx \Delta {\mathpzc g}_p(x;\zeta)$ to $x_0=0$.
The result is somewhat larger than that obtained via Eq.\,\eqref{uvlimit4PolB} because, as noted above, flavour thresholds work to suppress the number of final states available to a splitting gluon, thereby preserving strength/support in glue DFs.

All results listed in Table~\ref{tablepolarised} for the anomaly corrected contribution of quark helicities to the proton spin are depicted in Fig.\,\ref{Figa0E} and compared with the value in Ref.\,\cite{COMPASS:2016jwv}-[COMPASS].
Evidently, Eqs.\,\eqref{uvlimit4Pol}, \eqref{Eqa0E} and Eqs.\,\eqref{Eqa0E}, \eqref{PolGlueThresh}, deliver agreement with the COMPASS extraction,  
irrespective of the sources used for $\langle x \rangle_{\Delta\Sigma_{p}}^{\zeta}$, $\langle x \rangle_{{\mathpzc V}_p}^\zeta$.
However, best agreement is found when using values produced by the parameter-free CSM analyses in Refs.\,\cite{Lu:2022cjx, ChenChen:2022qpy}.

\begin{figure}[t]
\centerline{%
\includegraphics[clip, width=0.44\textwidth]{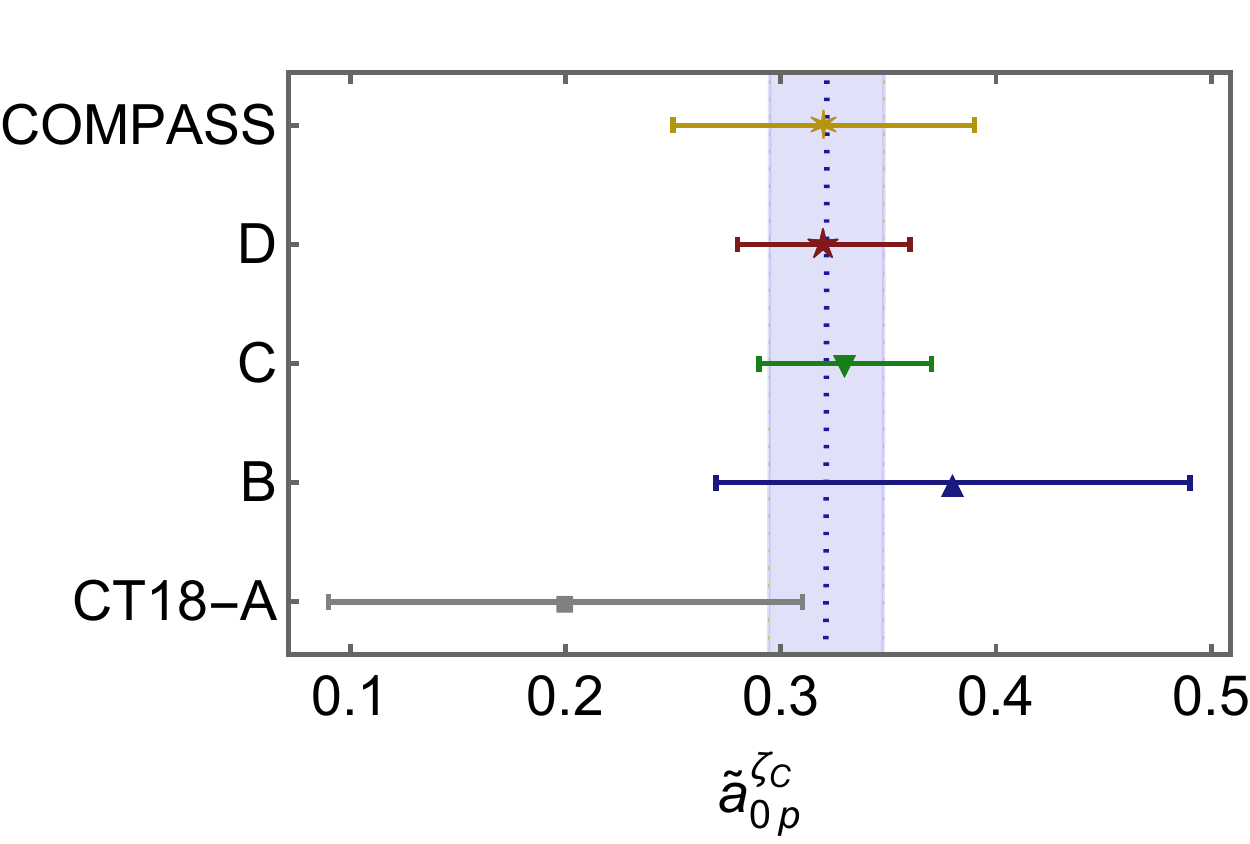}}
\vspace*{-2ex}

\caption{\label{Figa0E}
$\tilde a_{0p}^{\zeta = \zeta_{\rm C}}$, Eq.\,\eqref{Eqa0E}: non-Abelian anomaly corrected quark helicity contribution to the proton spin.
Data: gold asterisk and error bar -- Ref.\,\cite{COMPASS:2016jwv}-[COMPASS].
Theory: results drawn from Table~\ref{tablepolarised}.
Vertical dotted blue line within light-coloured band -- error weighted theory average: $0.32(3)$.
}
\end{figure}

\medskip

\noindent\emph{5.$\;$Summary}.
Every meson and baryon (hadron) is defined by its valence degrees of freedom.
Yet, within quantum field theory, such hadrons may be seen as containing infinitely many additional partons.
The associated distribution functions (DFs) are amongst the most important expressions of a hadron's structural characteristics.

Such DFs depend on the scale whereat the hadron's structure is resolved, \emph{i.e}., on the probe's wavelength.
In this connection, we identified $\zeta_{\cal H}$ -- the hadron scale -- as that value of the resolving scale at which valence degrees of freedom carry all properties of every hadron
and
introduced an effective charge, $\alpha_{1\ell}$, which, when used to integrate the leading-order scale evolution equations, delivers an evolution scheme for all DFs -- both unpolarised
and polarised -- that is all-orders exact.
Building solely upon these propositions, we presented an algebraic scheme which, given the valence DFs for any hadron at $\zeta_{\cal H}$, delivers predictions for all its DFs at $\zeta > \zeta_{\cal H}$.

The all-orders scheme delivers results that are typically independent of both the value of $\zeta_{\cal H}$ and the pointwise form of $\alpha_{1\ell}$; hence, it has real predictive power.
For instance, with Pauli blocking sufficient to explain the measured asymmetry of antimatter in the proton, one finds that the strange quark ($s+\bar s$) sea in the pion exceeds the $u+\bar u$ and $d+\bar d$ components and predicts qualitatively equivalent outcomes in other hadrons.

Furthermore, the scheme delivers a model-independent algebraic identity that relates the integrated strength of the in-proton gluon helicity DF, $\Delta G_p^\zeta$, to the analogous strength of the singlet polarised quark DF and the valence quark light-front momentum fraction, $\langle x \rangle_{{\mathpzc V}_p}^\zeta$.  It entails that $\Delta G_p^\zeta$ grows faster than $1/[\langle x \rangle_{{\mathpzc V}_p}^\zeta]^2$ as the valence momentum fraction decreases.
Exploiting this identity, contemporary measurements of the quark helicity contribution to the proton spin are reconciled with theory predictions.

%
\medskip
\noindent\emph{Acknowledgments}.
%
%
Work supported by:
National Natural Science Foundation of China (grant no.\,12135007);
Nanjing University of Posts and Telecommunications Science Foundation (grant no.\ NY221100);
Natural Science Foundation of Jiangsu Province (grant no.\ BK20220122);
Spanish Ministry of Science and Innovation (MICINN grant no.\ PID2019-107844GB-C22);
and
Junta de Andaluc{\'{\i}}a (grant no.\ P18-FR-5057).


\end{document}